\documentclass[preprint,showpacs,preprintnumbers,nofootinbib,amsmath,amssymb]{revtex4}

\usepackage{graphicx}% Include figure files
\usepackage{dcolumn}% Align table columns on decimal point
\usepackage{bm}% bold math

\begin{document}
\title{Refined Glauber model versus Faddeev calculations and experimental data for $\bm{pd}$ spin observables}
\author{\firstname{M.~N.}~\surname{Platonova}}
\email{platonova@nucl-th.sinp.msu.ru}
\author{\firstname{V.~I.}~\surname{Kukulin}}
\email{kukulin@nucl-th.sinp.msu.ru}
\affiliation{Skobeltsyn Institute of Nuclear Physics, \\
Lomonosov Moscow State University, Moscow RU-119991, Russia}
\date{\today}
\begin{abstract}
Spin-dependent observables in intermediate-energy $pd$ elastic
scattering within the framework of the refined Glauber model are
considered. The improvements include an account of all ten $pp$
and $pn$ helicity amplitudes at respective energies constructed on
the basis of modern phase-shift analysis, accurate deuteron wave
functions taken from the modern $NN$ force model and account of
charge-exchange effects. Predictions of the refined diffraction
model for the differential cross section and analyzing powers are
compared with exact three-body Faddeev calculations and the recent
experimental data. An amazingly good agreement between the results
of both theoretical approaches as well as between the refined
Glauber model and experiment in a wide angular range not only for
differential cross section but also for vector and tensor
analyzing powers has been found in the first time. Possible
reasons for this agreement are discussed.
\end{abstract}
\maketitle
\newpage
\section{Introduction}
\label{intro} It is well known that the Glauber diffraction model
\cite{G0,G1} is a convenient and reliable tool for the analysis of
scattering of fast hadrons (nucleons) by nuclei. Based on the
eikonal and fixed-scatterer approximations, it was specially developed more
than 50 years ago for the high- and intermediate-energy
regions where no exact theoretical treatments were available. So,
the validity of Glauber model could be tested previously only by
comparing its results with the respective experimental data.
The unexpected success of such a simple model in describing the
hadron-nucleus and nucleus-nucleus scattering at forward angles
caused numerous studies of the accuracy and the range of validity
of the Glauber formulation, as well as many attempts of extension
of this range. Different refinements of the initial simple model
have been introduced since then, and they included
corrections for non-eikonal and relativistic effects, Fermi
motion, etc. The last (in time) substantial steps taken in this
direction can be found in Refs.~\cite{GUR,BAL,ABJ,BBJ}. However,
the comprehensive analysis of various corrections to the Glauber
model has revealed \cite{H2} that many important corrections to
the initial model seems tend to compensate strongly each other, so
that an incorporation of only one of them can even worsen the
results of the initial simple model. So, it turned out to be
highly nontrivial to improve the initial Glauber approach.

Another serious problem with this model seems is its rather
restricted range of applicability, i.e., it should work well, in
general, at sufficiently high energies and forward angles. However,
it would be extremely interesting (for many practical
applications) to know these limits more definitely, although they
are dependent upon the particular problem to be solved. Fortunately,
nowadays we can learn much more than before about these limits for
some important cases by comparing the predictions of the Glauber
model against the results of precise calculations within the framework
of the respective full models, i.e., without approximations
peculiar to the diffraction model. Among these cases allowing the
careful comparison with a numerically accurate treatment is the $Nd$
intermediate-energy scattering within a realistic three-body
model. Now we have a very nice opportunity to examine the accuracy
of the Glauber model by direct comparison of its predictions with
exact three-nucleon Faddeev calculations~\cite{GLO} which account for the
same (nucleonic) degrees of freedom and the same input on-shell
$NN$ amplitudes. Such a test will show qualitatively or even
quantitatively the validity of different approximations involved
in the Glauber model. To obtain fully realistic
conclusions the Glauber model itself must as realistic as
possible; i.e., it should include all fully realistic input
spin-dependent $NN$ amplitudes and all components of the target
(e.g., deuteron) wave function. Such generalization of the initial
model enables us to analyze the spin observables (which should be
much more sensitive to fine interference effects and different
approximations) as well as the unpolarized cross sections, so we
will be able to draw more quantitative and well-grounded
conclusions about the validity of the Glauber formulation. For a
meaningful comparison with exact three-body calculations, the
inputs of the model, i.e., $NN$ amplitudes and deuteron wave
functions, must also be the most accurate and coincide
with those used in the current Faddeev calculations. Because the
diffraction model includes on-shell $NN$ amplitudes only, they can
be taken from the experiment. Or, more definitely, one can take
these amplitudes from modern phase-shift analysis (PSA), so that
they will be on-shell equivalent to those derived from realistic
$NN$ potential models entering the Faddeev equations (in, of course,
the energy region where such potentials describe accurately the
$NN$ experimental data).

The fully realistic Faddeev equations for $Nd$ scattering have
been solved up to now only for the incident energies below 350 MeV
in the laboratory frame~\cite{FC250,EXP248}.\footnote{There is only a single full three-body
calculation \cite{EXPAD4} for $pd$ scattering at the proton
incident energy $T_p \simeq 400$ MeV, but its results are still
preliminary and have not been published yet.} Complications which
arise with growing energy are connected with limitations of highly
precise $NN$ potentials involved as well as with hard
computational problems. Recently \cite{LIN}, the Faddeev
calculations at higher energies (up to 2 GeV) have been carried
out, but only in a schematic model with three identical bosons
interacting through a scalar central potential of the Malfliet-Tjon
type. In this model, a detailed comparison with the Glauber
approach for total and differential elastic cross sections was
also performed \cite{ELSTER}.

In the present paper, we tested the validity of the Glauber model
with a fully realistic two-body input. First, we generalized
the initial model by incorporating the full spin dependence of
$NN$ amplitudes and high-quality deuteron wave function as well as
the charge-exchange effects. We analyzed the differential cross
sections and polarization observables in $pd$ elastic scattering
at the energies of a few hundred MeV, which seems already high
enough to apply the generalized Glauber approach but still low
enough to compare its predictions with those of exact realistic
Faddeev calculations. Moreover, it was demonstrated \cite{FC250}
that at such moderate energies, relativistic effects do not play a
significant role at small and medium scattering angles, so the
nonrelativistic treatment seems to be sufficient on such
conditions. To confirm our conclusions and to obtain a more
clear understanding of the phenomena in question, the comparison of
the results for both theoretical approaches, i.e., Glauber and
Faddeev, with available experimental data is also presented. From
all these comparisons, one can draw more definite conclusions
about the true range of validity of the refined Glauber model.

The content of the paper is as follows. In Sec.~\ref{model}, we generalize
the initial diffraction model by incorporating all ten $NN$
helicity amplitudes (five are for $pp$ and five are for $pn$
scattering) and develop a convenient Gaussian-like parametrization
of these amplitudes. Also we build a multi-Gaussian expansion for
realistic deuteron $S$- and $D$-wave functions. The convenient
analytical representation for main input ingredients of the model
makes it possible to derive all 12 invariant $pd$ amplitudes
in fully analytical forms. In Sec.~\ref{results}, we present the main results of
the work. The detailed and comprehensive discussion of the
obtained results and some physical arguments which can help to
interpret our findings more clearly are presented in Sec.~\ref{discuss}.
Sec.~\ref{sum} is devoted to formulation of the conclusions. Two
appendixes include some important details of the calculations
within the framework of the refined Glauber model. In
Appendix~\ref{A}, we present the explicit interrelations between
all $pd$ invariant amplitudes through $NN$ invariant amplitudes
and deuteron formfactors. In Appendix~\ref{B}, details of
analytical integration in the double-scattering terms of the $pd$
amplitudes are given.

\section{Refined Glauber model}
\label{model}

To explore high-precision spin-dependent
$NN$ interactions for describing $pd$ elastic scattering, the
conventional Glauber model and its basic formulas which relate
$pd$ amplitude to the input $NN$ amplitudes and the deuteron wave
function have to be generalized. In preceding years, some papers
have been published that considered the following contributions
separately: (i) spin dependence of $NN$ amplitudes \cite{KSP,FSP},
(ii) $D$ wave of the deuteron \cite{HD,GD}, (iii) isospin dependence
of $NN$ amplitudes, i.e., double charge-exchange contribution to
$pd$ elastic scattering \cite{WEX,GEX}. All these items were
included later in the so-called relativistic multiple-scattering
theory \cite{ABJ,BBJ} which went beyond the Glauber framework by
accounting for corrections to the eikonal and fixed-scatterer
approximations and some relativistic effects as well. It is well
known, at least qualitatively, that different corrections to the
Glauber model tend to cancel each other substantially
\cite{H2}, so it is hard to improve the Glauber model
essentially. Besides, the modified versions are much more
complicated than the initial model. So, we have generalized just
the initial Glauber formulation by including the above-mentioned
items without any further corrections to the diffraction model
itself, thus staying within the original Glauber framework.

\subsection{Definition of observables}
\label{defobs}
First of all, we need to define the differential
cross section and spin-dependent observables in terms of the $pd$
elastic-scattering amplitude. The differential cross section is
connected to the above amplitude $M$ by the relation\footnote{Our
normalization is different from the standard one by the
Lorentz-invariant factor ${8 \sqrt{\pi} I(s,m_p,m_d) \! \equiv \!
4 \sqrt{\pi [s\!-\!(m_p + m_d)^2][s\!-\!(m_p - m_d)^2]}}$, where
$s$ is the $pd$ invariant mass squared, and $m_p$ and $m_d$ are the
proton and deuteron masses. Such normalization is chosen in order
to simplify the final formulas.}
\begin{equation}
\label{dsm}
 d\sigma\! / \!dt = {\textstyle
\frac{1}{6}}{\rm Sp}\,\bigl(MM^+\bigr),
\end{equation}
where $t = -q^2$ is the momentum transfer
squared.\footnote{Although we work in the laboratory frame
according to the initial Glauber suggestion, we should throughout
keep in mind the relation $t = -q^2$ for consistency. This
relation is valid in the center-of-mass frame and approximately
valid in the laboratory frame at small momentum transfers.
Physically, the difference between the variables $t$ in these two
frames originates from recoil effects which are neglected in the
Glauber formalism due to the fixed-scatterer approximation. So,
this difference should not be accounted for without careful
treatment of recoil effects as well as other corrections to the
Glauber model which all become significant at large momentum
transfers.} As for spin-dependent observables, in this work we
concentrate mainly on the vector and tensor analyzing powers. For
the proton and deuteron vector analyzing powers ($A_{\alpha}^p$
and $A_{\alpha}^d$) and for the deuteron tensor analyzing powers
($A_{\alpha \beta}$) we take the standard formulas
\begin{equation*}
 A_{\alpha}^p = {\rm Sp}\,\bigl(M \sigma_{\alpha}M^+\bigr)/{\rm
Sp}\,\bigl(MM^+\bigr), \ \ \ A_{\alpha}^d = {\rm Sp}\,\bigl(M
S_{\alpha}M^+\bigr)/{\rm Sp}\,\bigl(MM^+\bigr),
\end{equation*}
\begin{equation}
\label{apm} A_{\alpha \beta} = {\rm Sp}\,\bigl(M S_{\alpha \beta}
M^+\bigr)/{\rm Sp}\,\bigl(MM^+\bigr),
\end{equation}
where $\frac{1}{2}\sigma_{\alpha}$ and
$S_{\alpha}=\frac{1}{2}(\sigma_{n\alpha} + \sigma_{p\alpha})$ are
the spin matrices of the proton and deuteron, $S_{\alpha \beta}=
\frac{3}{2}(S_{\alpha}S_{\beta} + S_{\beta}S_{\alpha}) -
2\delta_{\alpha \beta}$ is a quadrupole operator, and $\alpha,\beta
\in \{x,y,z\}$.

The total amplitude $M$ can be expanded on the amplitudes
invariant under space rotations and space-time reflections. For
the $pd$ case, there are 12 such invariant amplitudes
$A_1$--$A_{12}$, and the amplitude $M$ (in nonrelativistic
formulation) is expressed through them as
\begin{eqnarray}
\label{ma}
   M[{\bf p},{\bf  q}; {\bm \sigma}, {\bf  S}] &=& \bigl(A_1 + A_2 \,{\bm
\sigma}\hat{n}\bigr) + \bigl(A_3 + A_4 \,{\bm
\sigma}\hat{n}\bigr)({\bf S}\hat{q})^2 + \bigl(A_5 + A_6 \,{\bm
\sigma}\hat{n}\bigr)({\bf S}\hat{n})^2 + \nonumber \\
 & & + A_7 \, ({\bm \sigma}\hat{k})({\bf S}
\hat{k}) + A_{8} \,{\bm \sigma}\hat{q} \bigl(({\bf S}\hat{q})({\bf
S}\hat{n}) \!+\! ({\bf S}\hat{n})({\bf S}\hat{q})\bigr) +
\bigl(A_9 + A_{10} \,{\bm
\sigma}\hat{n}\bigr){\bf S}\hat{n} + \nonumber \\
 & & + A_{11} \,({\bm \sigma}\hat{q})({\bf  S}\hat{q}) + A_{12} \,{\bm \sigma}\hat{k} \bigl(({\bf S}
\hat{k})({\bf S}\hat{n}) \!+\! ({\bf S} \hat{n})({\bf
S}\hat{k})\bigr),
\end{eqnarray}
where the unit vectors $\hat{k} = ({\bf p}+{\bf p'})/|{\bf p}+{\bf
p'}|, \ \ \hat{q} = ({\bf p}-{\bf p'})/|{\bf p}-{\bf p'}|, \ \
\hat{n} = \hat{k} \times \hat{q}$ and ${\bf p}$, ${\bf p'}$ are
the momenta of the incident and outgoing proton respectively.

Now all the $pd$ observables can be written in terms of invariant
amplitudes $A_1$--$A_{12}$. Defining the directions of coordinate
axes $\hat{e}_x = \hat{q}, \,\, \hat{e}_y = \hat{n}, \,\,
\hat{e}_z = \hat{k}$ and applying the standard trace technique, one
gets for the differential cross section and nonvanishing
analyzing powers the following expressions:
\begin{eqnarray}
\label{apa}
  d\sigma\! / \!dt \!\!\! &=& \!\!\! |A_1|^2 + |A_2|^2 + {\textstyle
\frac{2}{3}}\Bigl(\sum\limits_{i=3}^{12}{|A_i|^2} + {\rm
Re}\,\bigl[2A_1^*(A_3 + A_5) + 2A_2^*(A_4 + A_6) + A_3^*A_5 +
A_4^*A_6\bigr]\Bigr), \nonumber \\
  {A_y}^p \!\!\! &=& \!\!\! 2\,{\rm Re}\,\bigl[2\,(A_1^* + A_3^*
+A_5^*)(A_2 + A_4 + A_6) + A_1^*A_2 - A_3^*A_6 - A_4^*A_5 +
2A_9^*A_{10}\bigr]/(3\,d\sigma\! / \!dt), \nonumber \\
  {A_y}^d \!\!\! &=& \!\!\! 2\,{\rm Re}\,\bigl[(2A_1^* + A_3^*
+2A_5^*)A_9 + (2A_2^* + A_4^* + 2A_6^*)A_{10} + A_7^*A_{12} +
A_8^*A_{11}\bigr]/(3\,d\sigma\! / \!dt), \nonumber \\
  A_{yy} \!\!\! &=& \!\!\! \Bigl(2\,\bigl(|A_5|^2 + |A_6|^2 +
|A_9|^2 + |A_{10}|^2\bigr) - \bigl(|A_3|^2 + |A_4|^2 + |A_7|^2 +
|A_8|^2 +
|A_{11}|^2 + |A_{12}|^2\bigr) + \nonumber \\
  & & + 2\,{\rm Re}\,\bigl[A_1^*(2A_5 - A_3) + A_2^*(2A_6 - A_4) + A_3^*A_5 + A_4^*A_6\bigr]\Bigr)/(3\,d\sigma\! / \!dt), \nonumber \\
  A_{xx} \!\!\! &=& \!\!\! \Bigl(2\,\bigl(|A_3|^2 + |A_4|^2 +
|A_{11}|^2 + |A_{12}|^2\bigr) - \bigl(|A_5|^2 + |A_6|^2 + |A_7|^2
+ |A_8|^2
+|A_{9}|^2 + |A_{10}|^2\bigr) + \nonumber \\
  & & + 2\,{\rm Re}\,\bigl[A_1^*(2A_3 - A_5) + A_2^*(2A_4 - A_6) + A_3^*A_5 + A_4^*A_6\bigr]\Bigr)/(3\,d\sigma\! / \!dt), \nonumber \\
  A_{zz} \!\!\! &=& \!\!\! -A_{yy} -A_{xx}, \nonumber \\
  A_{xz} \!\!\! &=& \!\!\! {\rm Im}\,\bigl[A_3^*A_9 + A_4^*A_{10} - A_7^*A_{12} -
A_8^*A_{11}\bigr]/(d\sigma\! / \!dt).
\end{eqnarray}

\subsection{Generalization of initial Glauber formalism}
\label{gengl}
In the initial Glauber model, the $pd$ scattering
amplitude as the function of transferred momentum ${\bf q}$ is
represented as a sum of two terms corresponding to single and
double scatterings of the incident proton off target nucleons:
\begin{equation}
\label{msd} M({\bf q}) = M^{(s)}({\bf q}) + M^{(d)}({\bf q}).
\end{equation}
With the use of eikonal and fixed-scatterer approximations, the
single- and double-scattering amplitudes are expressed in terms of
the on-shell $NN$ amplitudes ($pp$ amplitude $M_p$ and $pn$
amplitude $M_n$) and the deuteron wave function $\Psi_d$ as
\begin{eqnarray}
\label{msmd}
 M^{(s)}({\bf q}) &=& {\textstyle \int} d^{3}r e^{i{\bf q}{\bf r}/2}
\Psi_d({\bf r}) \bigl[M_n({\bf q}) + M_p({\bf q})\bigr]
 \Psi_d({\bf r}), \nonumber \\
 M^{(d)}({\bf q}) &=& {\textstyle \frac{i}{4 \pi^{3/2}}\int} d^{2}q'{\textstyle\int} d^{3}r e^{i{\bf
q'}{\bf r}} \Psi_d({\bf r}) \bigl[M_n({\bf q_2}) M_p({\bf q_1})
 + M_p({\bf q_1}) M_n({\bf q_2})\bigr] \Psi_d({\bf r}),
\end{eqnarray}
where the vectors ${\bf q_1} = {\bf q}/2 - {\bf q'} , \ \ {\bf
q_2} = {\bf q}/2 + {\bf q'}$ have been introduced for momenta
transferred in collisions with individual target
nucleons.\footnote{In the expression~(\ref{msmd}) for the amplitude
$M_d$, we have omitted the term arising from the commutator of the
amplitudes $M_n({\bf q_2})$ and $M_p({\bf q_1})$ \cite{GEX}. This
term gives only a small contribution to the intermediate-energy
$pd$ elastic scattering due to relative smallness of
spin-dependent $NN$ amplitudes and the deuteron $D$ wave.}

The double-charge-exchange process contributes to elastic
scattering as well. This contribution is significant at incident
energies $T_p \lesssim 1$ GeV, so we should include it in the
model. It was already done in Ref.~\cite{GEX} by incorporating the
isospin structure of the general $NN$ amplitude and averaging over
the isoscalar deuteron ground state. This operation leads to an
additional term in double-scattering amplitude
\begin{equation}
\label{mc}  M^{(c)}({\bf q}) = -{\textstyle \frac{i}{4
\pi^{3/2}}\int} d^{2}q'{\textstyle\int} d^{3}r e^{i{\bf q'}{\bf
r}} \Psi_d({\bf r}) \bigl[M_n({\bf q_2}) - M_p({\bf
q_2})\bigr]\bigl[M_n({\bf q_1}) - M_p({\bf q_1})\bigr] \Psi_d({\bf
r}).
\end{equation}
The neglect of the spin dependence in $NN$ amplitudes and deuteron
wave function reduces Eqs.~(\ref{msd}) and (\ref{msmd}) to the
conventional Glauber formulas. Furthermore, the
double-charge-exchange amplitude $M_c$ vanishes in a widely used
approximation $M_n = M_p$ (it corresponds to neglecting the
isospin dependence of the general $NN$ amplitude). In the
realistic case, with which we are here concerned, the accurate
incorporation of both spin and isospin degrees of freedom is
required. While the latter is done simply by adding the term $M_c$
to the double-scattering amplitude $M_d$, the inclusion of spin
structure of both $NN$ amplitudes and deuteron wave function in the
Glauber model is much more involved. We take $NN$ amplitudes in
the form
\begin{eqnarray}
\label{mi}
 M_i[{\bf p}, {\bf q};{\bm\sigma},{\bm\sigma}_i] &=&
A_i + C_i\,{\bm\sigma}\hat{n} + C'_i\,{\bm\sigma}_i\hat{n}+
B_i\,({\bm\sigma}\hat{k})({\bm\sigma}_i\hat{k}) + \nonumber \\
 & & + \, \bigl(G_i+H_i\bigr)({\bm\sigma}\hat{q})({\bm\sigma}_i\hat{q})
+ \bigl(G_i-H_i\bigr)({\bm\sigma}\hat{n})({\bm\sigma}_i\hat{n}),
\end{eqnarray}
where $i=n,p$. In the laboratory frame, one should
distinguish the amplitudes $C$ and $C'$.

For the deuteron wave function, we use the standard expression
\begin{equation}
\label{psi}
 \Psi_d[{\bf r};{\bm\sigma}_n,{\bm\sigma}_p] = {\textstyle
\frac{1}{\sqrt{4\pi}r}}\bigl(u(r) + {\textstyle
\frac{1}{2\sqrt{2}}}w(r)\,
S_{12}[\hat{r};{\bm\sigma}_n,{\bm\sigma}_p]\bigr),
\end{equation}
where $u$ and $w$ are the radial wave functions for $S$ and
$D$ waves, and $S_{12}[\hat{n};{\bf v}_1,{\bf v}_2] = 3({\bf
v}_1\hat{n})({\bf v}_2\hat{n}) - ({\bf v}_1{\bf v}_2)$.

After substituting expressions~(\ref{mi}) and~(\ref{psi}) into Eqs.~(\ref{msmd}) and (\ref{mc}), and making some spin algebra with noncommuting operators $M_n,
M_p$ and $\Psi_d$, one gets rather complicated general formulas for the
$pd$ amplitudes $M^{(s)}$, $M^{(d)}$, and $M^{(c)}$ expressed through the input $NN$
amplitudes $A_i,B_i,C_i,C'_i,G_i,H_i$ ($i=n,p$) and the deuteron form
factors $S_0^{(0)},S_0^{(2)},S_2^{(1)},S_2^{(2)}$. To simplify
further derivation, one can employ the
smallness of the spin-dependent $NN$ amplitudes (say, $\mathcal{B}_i$)
compared to spin-independent ones ($A_i$) at high energies as well
as the smallness of the deuteron $D$-wave $w$ compared to $S$-wave $u$~\cite{PKYAF}.
So, the terms containing products
$\mathcal{B}_i^k w^l$ with $k+l \geqslant 3$ can be dropped out of the expressions for
the amplitudes $M^{(s)}$, $M^{(d)}$, and $M^{(c)}$ on
definite conditions. In fact, the ratio of spin-dependent amplitudes
$\mathcal{B}_i$ to spin-independent ones $A_i$ is strongly
decreasing when the energy rises, so that such an approximation in the $pd$ amplitudes, being
quite accurate at intermediate energies $T_p \sim 1$ GeV, can be
unsatisfied at lower energies $T_p \sim 100$ MeV. This observation has
nothing to do with the validity of the Glauber model itself at such
lower energies, and it should be kept in mind when doing the
careful comparison between the present version of the Glauber
model and experimental data for spin analyzing powers (especially
for tensor ones which are more sensitive to fine spin-dependent
effects) in Sec.~\ref{results}.

After the above simplification, Eqs.~(\ref{msmd}) and (\ref{mc}) can be easily
integrated over $d^3r$. In doing this, we make use of the deuteron
form factor, which is defined as
\begin{eqnarray}
\label{s}
 S[{\bf q};{\bm\sigma}_n,{\bm\sigma}_p] &=& {\textstyle \int} d^3 r \, e^{i{\bf q}{\bf
r}}\,|\Psi_d[{\bf r};{\bm\sigma}_n,{\bm\sigma}_p]|^2 =
\nonumber \\
&=& S_0(q) - {\textstyle\frac{1}{2\sqrt{2}}} S_2(q) \,
S_{12}[\hat{q};{\bm\sigma}_n,{\bm\sigma}_p].
\end{eqnarray}
It is convenient to divide the monopole and quadrupole form
factors, $S_0$ and $S_2$, into two parts which correspond to
different multiplicities of the $D$-wave function $w$, i.e.,
\begin{equation}
\label{s0s2} S_0(q) = S_0^{(0)}(q) + S_0^{(2)}(q),  \ \ \ S_2(q) =
S_2^{(1)}(q) + S_2^{(2)}(q),
\end{equation}
where
\begin{eqnarray}
\label{s00}
 S_0^{(0)}(q) &=& {\textstyle \int\limits_0^{\infty}}dr \,
u^2(r)\, j_0(qr), \qquad \quad
S_0^{(2)}(q) =  {\textstyle \int\limits_0^{\infty}}dr \, w^2(r)\, j_0(qr), \nonumber \\
S_2^{(1)}(q) &=& 2 {\textstyle \int\limits_0^{\infty}}dr \,
u(r)w(r)\, j_2(qr), \ \ \
S_2^{(2)}(q) = - 2^{-1/2} {\textstyle
\int\limits_0^{\infty}}dr \, w^2(r)\, j_2(qr).
\end{eqnarray}

Eventually, using the expansion~(\ref{ma}) for the total $pd$ amplitude $M$,
one obtains the explicit interrelations between all 12
invariant $pd$ amplitudes and 12 invariant input $NN$
amplitudes and also different components of the deuteron form
factor (for the final formulas and details of analytic ${\bf
q'}$ integration in the double-scattering amplitudes, see
Appendixes~\ref{A} and \ref{B}, respectively). Having these
interrelations and proper two-body input in hand, one can
calculate straightforwardly the $pd$ differential cross section
and all polarization observables on the basis of the refined
Glauber model.

\subsection{Parametrization of the $\bm{NN}$ amplitudes and deuteron wave function}
\label{param}

The Glauber model deals with $pd$ and $NN$
amplitudes defined in the laboratory frame. However, it is more
convenient to treat the $NN$ helicity amplitudes in the
two-nucleon center-of-mass frame. It is easy to show that the
laboratory amplitudes $A,B,C,G,H$ at small $q$ can be
straightforwardly expressed through the conventional helicity
amplitudes $N_0,N_1,N_2,U_0,U_2$ (or $\phi_1$--$\phi_5$) as
\begin{gather}
 A \approx N_0 = (\phi_3 +
 \phi_1)/2, \ \ B \approx -U_0 = (\phi_3 -
 \phi_1)/2, \nonumber \\
 C \approx iN_1 = i\phi_5, \nonumber \\
 G \approx (U_2 - N_2)/2 = \phi_2/2, \ \ H \approx (U_2 + N_2)/2
= \phi_4/2.
\label{anu}
\end{gather}
Here, in making appropriate approximations we do not go beyond the
diffraction model. It was also demonstrated \cite{SOR} that the
amplitude $C'$ (see Eq.~(\ref{mi})) in high-energy small-angle
limit is distinguished only by a relativistic correction from the
amplitude $C$, i.e.
\begin{equation}
\label{c'}
 C' \approx C + (q/2m)N_0.
\end{equation}
Moreover, both the amplitudes, $C$ and $C'$, are small at high
energies in comparison to the other amplitudes, so that the above
correction is hardly playing a significant role, but it still
should be included for consistency.

\begin{figure}
\begin{center}
\resizebox{0.7\columnwidth}{!}{\includegraphics{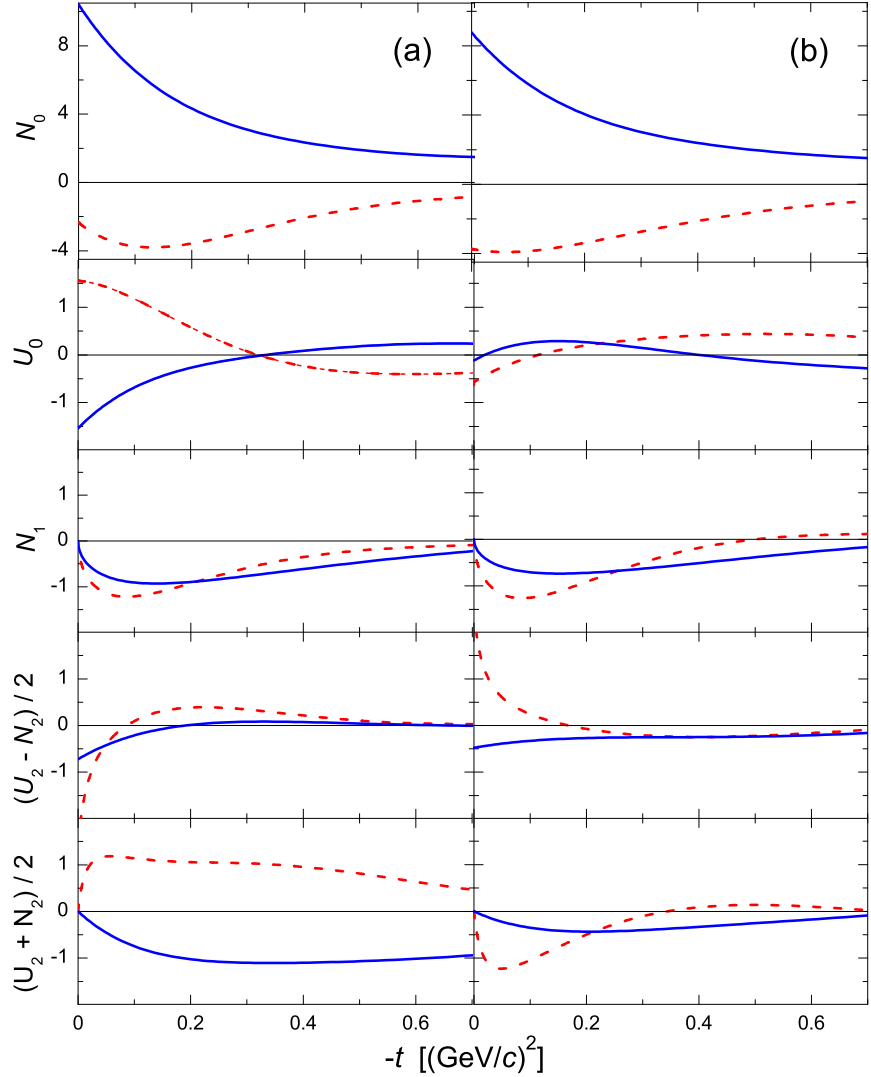}}
\end{center}
\caption{Combinations of the $NN$ helicity amplitudes (in units
$\sqrt{\rm mb}$/GeV), which correspond to the laboratory $NN$ amplitudes
used in our calculations (see Eq.~(\ref{anu})). The $pp$ amplitudes
are shown in column (a), the $pn$ amplitudes are given in column (b).
The dashed lines correspond to the real parts of the amplitudes,
while the solid lines represent their imaginary parts.} \label{nnamp}
\end{figure}

All the helicity $pp$ and $pn$ amplitudes at the energy $T_p = 1$
GeV are displayed in Fig.~\ref{nnamp}. These amplitudes are
built in the present work on the basis of recent PSA
\cite{SAID}, and we used a special code \cite{PANN} to reconstruct the
$pp$ and $pn$ helicity amplitudes from the PSA data. As is clearly
seen from Fig.~\ref{nnamp}, the amplitude $N_0$ superiors
significantly all the other helicity amplitudes. It is also
clearly seen that the corresponding $pp$ and $pn$ amplitudes are
distinguished from each other significantly while in early works
on the diffraction approach they have been chosen to be the same
for the sake of simplicity.

To parametrize the $NN$ helicity amplitudes, it is very convenient
to employ a Gaussian series representation with an explicit
separation of the behavior near $q = 0$:
\begin{gather}
 N_0(q) = {\textstyle
\sum\limits_{j=1}^n}\, C_{a,j}\exp(-A_{a,j}\,q^2), \ \ \ U_0(q) =
{\textstyle \sum\limits_{j=1}^n}\, C_{b,j}\exp(-A_{b,j}\,q^2),
\nonumber \\
 N_1(q) = q\,{\textstyle \sum\limits_{j=1}^n}\,
C_{c,j}\exp(-A_{c,j}\,q^2), \nonumber \\
(U_2(q)-N_2(q))/2 = {\textstyle \sum\limits_{j=1}^n}\,
C_{g,j}\exp(-A_{g,j}\,q^2), \nonumber \\
(U_2(q)+N_2(q))/2 = q^2\, {\textstyle \sum\limits_{j=1}^n}\,
C_{h,j}\exp(-A_{h,j}\,q^2).
\label{nug}
\end{gather}
Here the subscripts $a,b,c,g,h$ in the parameters $C,A$ denote the
respective laboratory $NN$ amplitudes (see
Eq.~(\ref{anu})).\footnote{In explicit calculations we explored two
different relations (and two sets of parameters $C,A$) for each
helicity amplitude, i.e., one for its real part and one for the
imaginary part. Here just the general forms which fit both real
and imaginary parts of the amplitudes are given for simplicity.}
In our calculations we took $n = 5$, i.e. five Gaussian terms in
all above sums. With this choice, we found that the Gaussian
approximated $NN$ amplitudes are very near to the exact ones in
the forward hemisphere~\cite{PKYAF}. The visible deviations begin only at large
angles where the Glauber model demands a fast vanishing of all the
underlying amplitudes. The rise in magnitude of the true $pp$
helicity amplitudes is due to Pauli principle, according to which
the whole $pp$ amplitude must be antisymmetrized. This
antisymmetrization is essential in large-angle $pd$ scattering
only through one-nucleon exchange mechanism, so that, the
diffraction model being derived for forward-angle scattering does
not account for this exchange mechanism. On the other hand, the
charge-exchange process which is responsible for the rising of
$np$ helicity amplitudes at large angles can contribute to $pd$
elastic scattering already at rather forward angles through the
double charge exchange, and thus, the latter mechanism is included
to our formalism explicitly.

For the deuteron wave function we explored the high-precise $NN$
potential model CD-Bonn~\cite{CDB}. To parametrize $S$- and $D$-wave
components of the function we have also employed the Gaussian
representation (with an additional factor $r^n$ to reproduce the
behavior near the origin):
\begin{equation}
\label{uwg} u(r) = r \, {\textstyle \sum\limits_{j=1}^m}\,
C0_j\exp(-A0_j\,r^2), \ \ \ w(r) = r^3 \, {\textstyle
\sum\limits_{j=1}^m}\, C2_j\exp(-A2_j\,r^2).
\end{equation}
In our calculations we have chosen $m=5$. With this number of
terms the approximated deuteron wave functions coincide with high
accuracy with the exact ones from the origin up to large distances
($r_{NN} \simeq 20$ fm). With the above parametrization of the
deuteron radial wave functions the form factors defined in
Eq.~(\ref{s00}) take the forms
\begin{eqnarray}
\label{sg}
 S_0^{(0)}(q) &=& {\textstyle \sum\limits_{i,j=1}^m}\,
C0_i C0_j
{\textstyle\frac{\sqrt{\pi}}{4\lambda_{00,ij}^{3/2}}}\exp(-x_{00,ij}), \nonumber \\
 S_0^{(2)}(q) &=& {\textstyle \sum\limits_{i,j=1}^m}\,
C2_i C2_j {\textstyle\frac{\sqrt{\pi}}{16\lambda_{22,ij}^{7/2}}}(4x_{22,ij}^2-20x_{22,ij}+15)\exp(-x_{22,ij}), \nonumber \\
S_2^{(1)}(q) &=& {\textstyle \sum\limits_{i,j=1}^m}\, C0_i C2_j
{\textstyle\frac{\sqrt{\pi}}{2\lambda_{02,ij}^{5/2}}}x_{02,ij}\exp(-x_{02,ij}), \nonumber \\
 S_2^{(2)}(q) &=& {\textstyle \sum\limits_{i,j=1}^m}\, C2_i
C2_j
{\textstyle\frac{\sqrt{2\pi}}{16\lambda_{22,ij}^{7/2}}}(2x_{22,ij}^2-7x_{22,ij})\exp(-x_{22,ij}),
\end{eqnarray}
where $\lambda_{kl,ij}=Ak_i+Al_j$, $x_{kl,ij} =
q^2/(4\lambda_{kl,ij})$, and $k,l = 0,2$.

\section{Results}
\label{results}

Using the above refined Glauber model we analyzed the $pd$
differential cross sections as well as proton and deuteron
analyzing powers at three intermediate energies: $T_p = 250$ and $440$
MeV and 1 GeV.\footnote{For the deuteron analyzing powers which are
measured in $dp$ scattering these are the equivalent proton
incident energies in the inverse kinematics, i.e., $T_p=T_d/2$.}
These energies were chosen because there is a considerable amount
of experimental data on $pd$ elastic observables in these energy
regions \cite{EXPAD4,EXP250,EXPAD2,EXPDS4,EXPAP392,EXPDS1,EXPAD1}.
Besides that, the two lower energies are appropriate to compare in
detail the predictions of our model with exact Faddeev results.

\begin{figure}
\begin{center}
\resizebox{0.5\columnwidth}{!}{\includegraphics{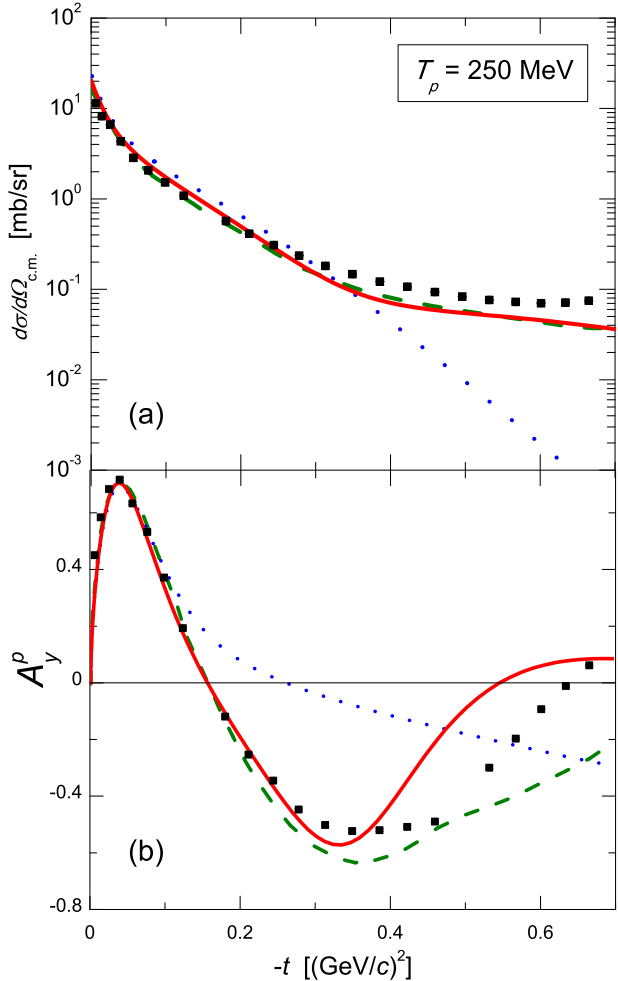}}
\end{center}
\caption{ Differential cross section (a) and proton analyzing power (b) in
$pd$ elastic scattering at the incident energy $T_p=250$ MeV. The
solid lines represent the results obtained within the refined
Glauber model, the dotted lines show the single-scattering
contribution only, while the dashed lines correspond to predictions of the exact Faddeev
calculations \cite{EXP250} with $NN$ potential CD-Bonn.  Experimental data (squares) are taken from
Ref.~\cite{EXP250}.} \label{ds-ay-250}
\end{figure}

\begin{figure}
\begin{center}
\resizebox{0.9\columnwidth}{!}{\includegraphics{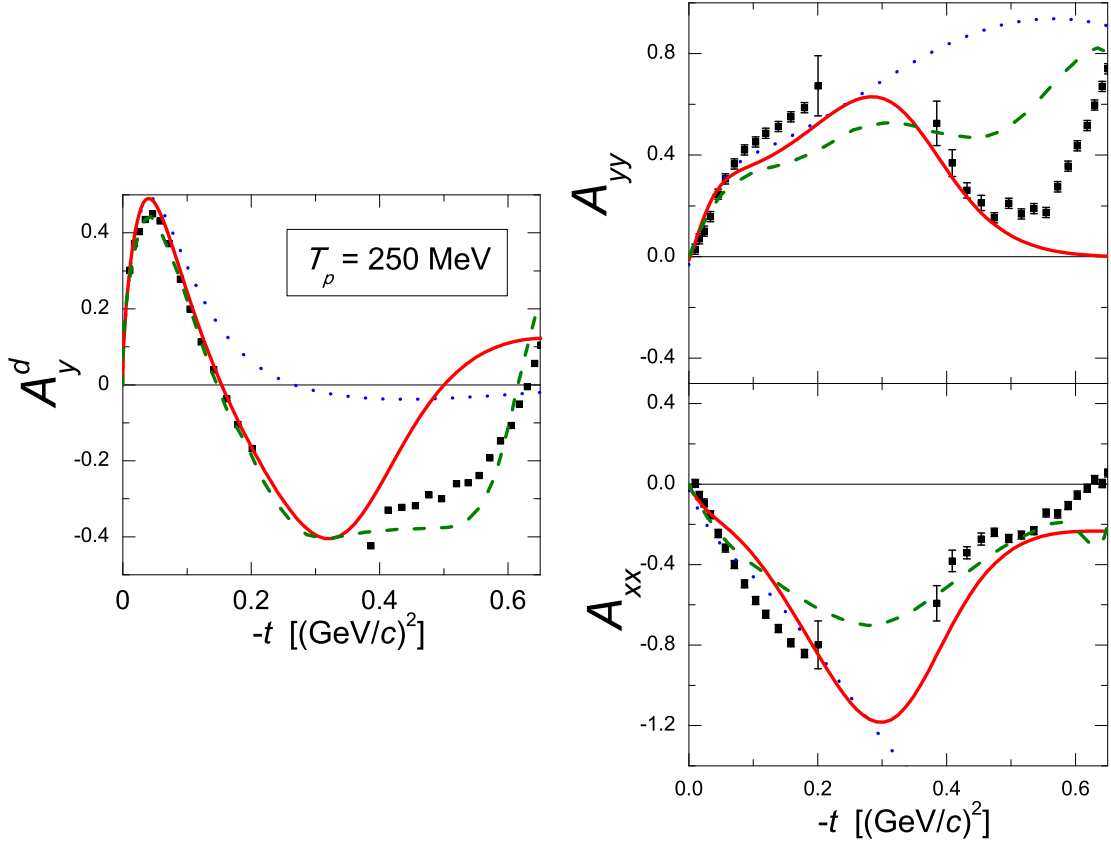}}
\end{center}
\caption{Deuteron vector (a) and tensor (b), (c) analyzing powers at the equivalent
proton energy $T_p=250$ MeV. For the notations, see
Fig.~\ref{ds-ay-250}. Results of the Faddeev calculations and
experimental data are taken from Ref.~\cite{EXPAD2} (for the energy $T_p=200$ MeV).}
\label{apd-250}
\end{figure}

\begin{figure}
\begin{center}
\resizebox{0.5\columnwidth}{!}{\includegraphics{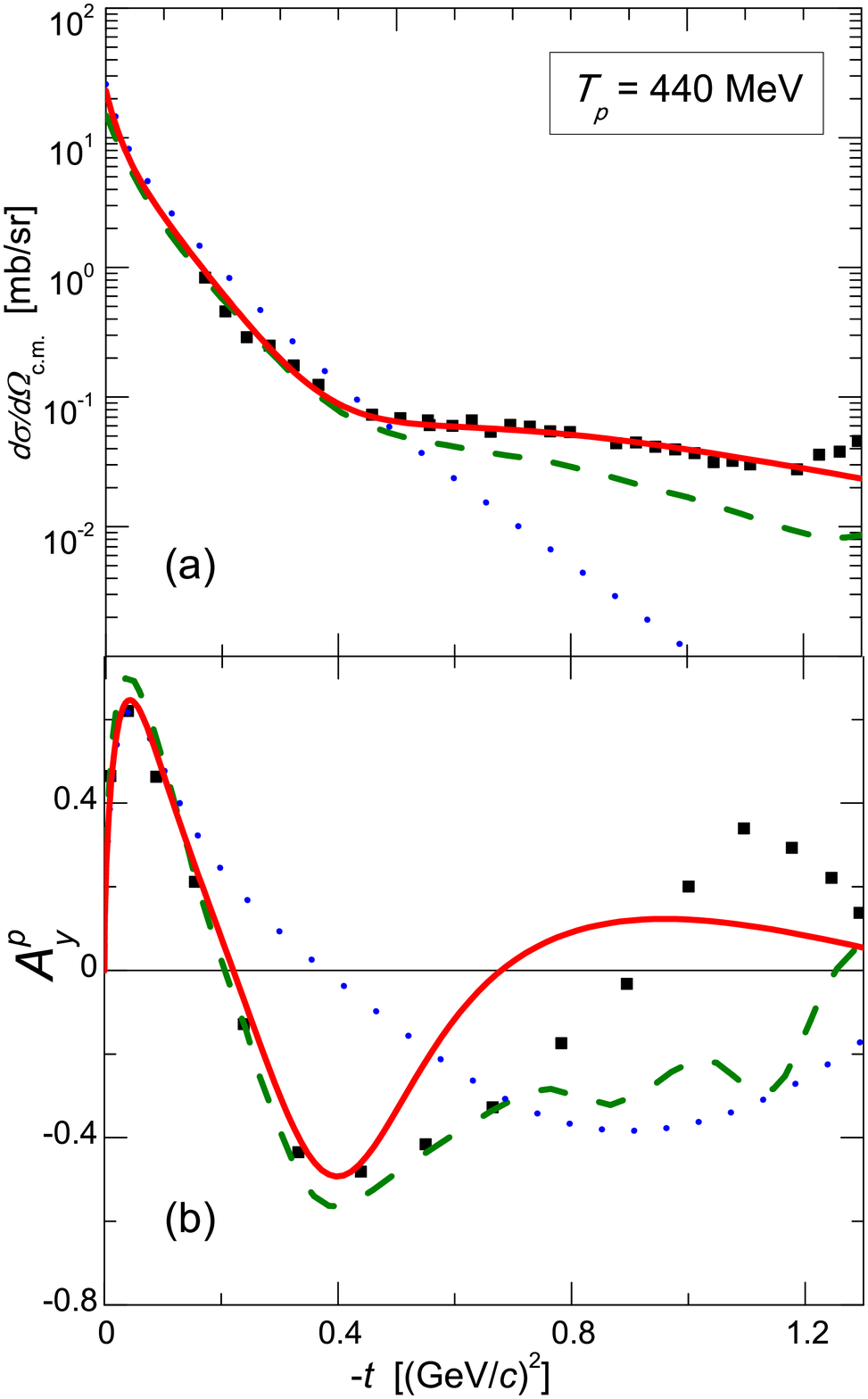}}
\end{center}
\caption{Differential cross section (a) and proton analyzing power (b) in
$pd$ elastic scattering at the incident energy $T_p=440$ MeV. For
the notations, see Fig.~\ref{ds-ay-250}. Results of Faddeev
calculations are taken from Refs.~\cite{EXPAD4} ($440$ MeV) and
\cite{EXPAP392} ($392$ MeV), experimental data --- from
Refs.~\cite{EXPDS4} ($425$ MeV) and \cite{EXPAP392} ($392$ MeV).}
\label{ds-ay-440}
\end{figure}

We start with the energy $T_p = 250$ MeV because the realistic
Faddeev calculations are well grounded for this energy. Results
for $pd$ differential cross section and proton analyzing power at
$T_p = 250$ MeV are represented in Fig.~\ref{ds-ay-250}. We
have also calculated the deuteron vector and tensor analyzing
powers at the equivalent proton energy $T_p = 250$ MeV. However
the exact Faddeev results and experimental data for these
observables are available in the literature just for a bit lower
energy $T_p = 200$ MeV. Our separate comparison between some
experimental data at $T_p = 200$ and $250$ MeV has shown that they
are very close to each other. So, our predictions at $T_p = 250$
MeV in comparison with exact three-body results and experimental
data at $T_p = 200$ MeV are displayed in Fig.~\ref{apd-250}.
In addition, we show the results of refined Glauber model at $T_p
= 440$ MeV (see Fig.~\ref{ds-ay-440}). The Faddeev calculations
with the fully realistic $NN$ interaction are not so reliable for
this energy, thus, we restrict ourselves with the differential
cross section and the proton analyzing power. We compared our
result for differential cross section at the energy $T_p = 440$
MeV with the result of Faddeev calculation at the same energy and
with experimental data at $T_p = 425$ MeV. For the comparison with
our result for proton analyzing power at $T_p = 440$ MeV, we took
existing (to date) Faddeev result and experimental data at a
bit lower energy $T_p = 392$ MeV.


\begin{figure}
\begin{center}
\resizebox{0.9\columnwidth}{!}{\includegraphics{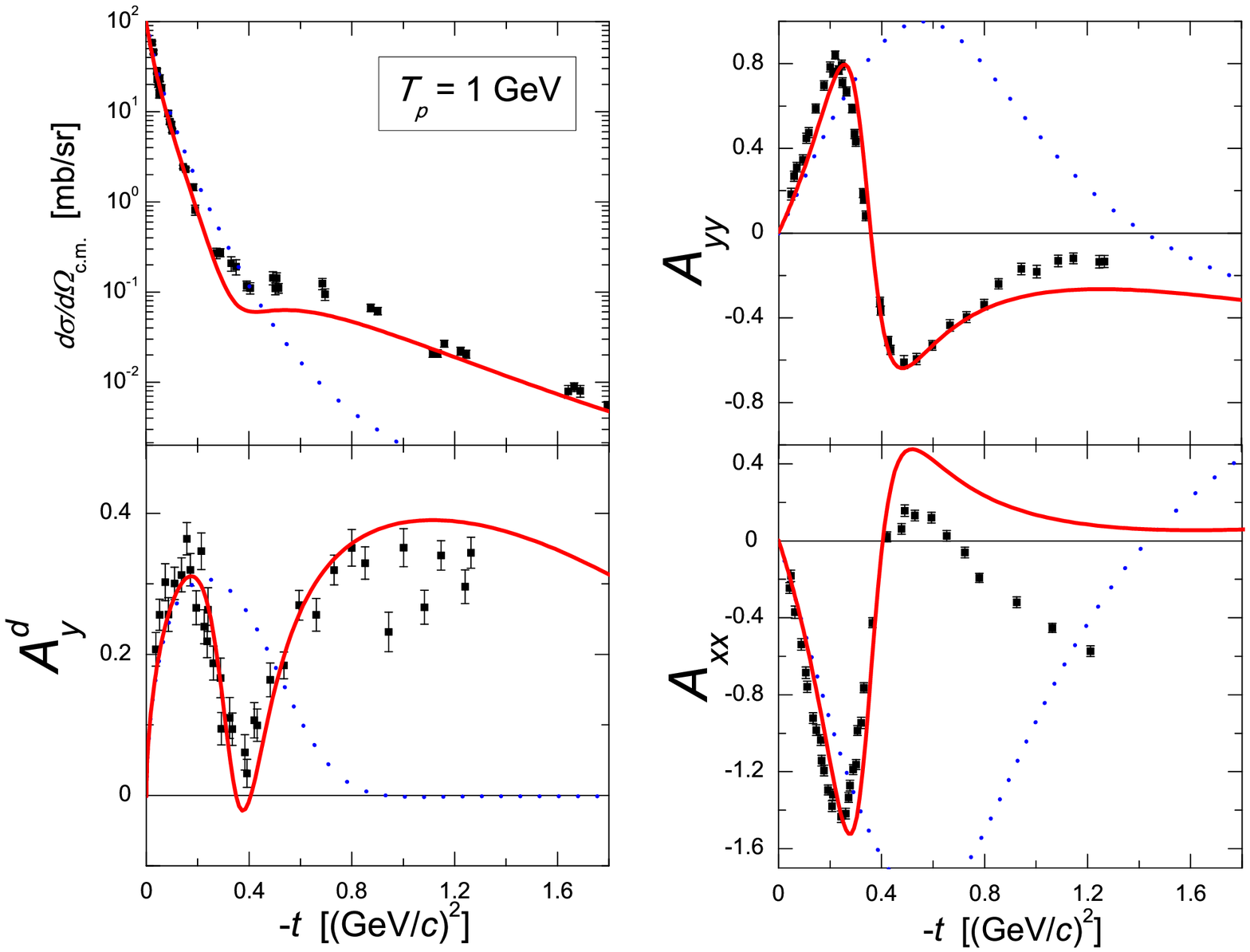}}
\end{center}
\caption{Differential cross section (a) and deuteron vector (b) and tensor (c), (d)
analyzing powers in $dp$ elastic scattering at
the equivalent proton energy $T_p=1$ GeV
calculated within the refined Glauber model. Dotted lines
show the contribution of single scattering only, solid lines
represent the full calculation. Experimental data (squares)
are taken from Refs.~\cite{EXPDS1} and \cite{EXPAD1}.}
\label{ds-apd-1000}
\end{figure}

Besides the comparison between the refined Glauber model
predictions and exact three-body Faddeev results, it would be
highly interesting to compare our results with existing
experimental data at the higher energy $T_p = 1$ GeV which is more
traditional for the diffraction model. This comparison has been
made for the differential cross section as well as for deuteron
vector and tensor analyzing powers. In Fig.~\ref{ds-apd-1000},
the predictions of our model together with respective experimental
data are displayed.

It is clearly seen from Figs.~\ref{ds-ay-250}--\ref{ds-apd-1000} that our results found
within the refined Glauber model are, in general, in a very
reasonable agreement with both exact three-body calculations and
experimental data at transferred momenta squared $|t| \lesssim
0.35$ (GeV/$c)^2$ for differential cross sections and vector and
tensor analyzing powers as well.\footnote{The agreement for tensor
analyzing powers at rather low energies ($T_p \simeq 250$ MeV) is
not so good as for differential cross sections and vector
analyzing powers (see Figs.~\ref{ds-ay-250} and \ref{apd-250}).
This fact is very likely related to our simplifying assumption
about relative smallness of the spin-dependent $NN$ amplitudes in
comparison to the large spin-independent ones (see the end of
Sec.~\ref{gengl}) and not to the validity of the Glauber
approximation itself.} This gives, at first glance, some
interesting deep puzzle because the good agreement with the exact
Faddeev calculations is seen in the region where the double
scattering (in Glauber model) dominates. However, instead of two
purely on-shell and no-recoil scatterings of incident proton by
two nucleons in deuteron within the Glauber model framework, the
Faddeev calculations include many fully off-shell rescatterings
with full account of recoil effects. We will discuss in detail
some possible physical reasons for such an amazing agreement in
the next section. Moreover, in
Figs.~\ref{ds-ay-250}--\ref{ds-ay-440} one can see some general new
trend: in those kinematical regions (at larger $|t|$) where the
refined diffraction model deviates essentially from the exact
Faddeev theory the exact $3N$ results begin to deviate also from
the experimental data. This gives another interesting
question to answer.

\section{Discussion}
\label{discuss}

It would be useful to arrange the general discussion of the
results obtained in this paper in a few separate points.

(i) Fully analytical formulas which relate all 12 invariant
$pd$ amplitudes to the accurate input $pp$ and $pn$ helicity
amplitudes (see Appendixes~\ref{A} and \ref{B}) allow us to not only
greatly simplify all the numerical calculations for $pd$ spin
observables but also to develop an efficient and convenient
algorithm for solving an important inverse scattering problem (at
fixed energy). This inverse problem can be formulated as follows:

(a) Having the precise intermediate-energy $nd$ spin observables and
differential cross section and by taking the respective $np$
helicity amplitudes at the same energy as a well-established input,
one can extract poorly known neutron-neutron scattering amplitudes
at the same energy.

(b) Or, alternatively, having in our possession the accurate $pd$
experimental data in the energy region $T_p > 1.1$ GeV, we can
find by inversion the proton-neutron scattering amplitudes which are
still poorly known at these energies.

Surely, a separate study should be done before doing this
inversion to establish here a real sensitivity of the input $pn$
amplitudes to the $pd$ cross sections and analyzing powers while
taking into consideration the experimental error corridor. So, such an
inversion opens a way to finding in principle the accurate $nn$
(or $pn$) scattering amplitudes from the precise $nd$ or $pd$
experimental data.

(ii) Our numerous calculations performed in this work on the basis
of refined Glauber model have been compared with the respective
exact Faddeev $3N$ calculations with mostly the same input
on-shell $NN$ amplitudes for differential cross section and vector
and tensor analyzing powers. For the numerous spin-dependent
observables, it was done for the first time. This direct comparison
has demonstrated clearly an amazingly good agreement between the
results of the refined diffraction model and exact $3N$
calculations, even at rather low energies $T_p \simeq 250$ MeV.
The agreement gets even more impressive when the collision energy
is rising. It should be stressed here that we observe this
nice agreement in the area where the double scattering in the
Glauber model approach becomes prevailing. This implies, among other
things, that the severe approximations made in the Glauber
approach just in the double-scattering treatment \cite{G1} really
work even at rather low energies.

\begin{figure}
\begin{center}
\resizebox{0.9\columnwidth}{!}{\includegraphics{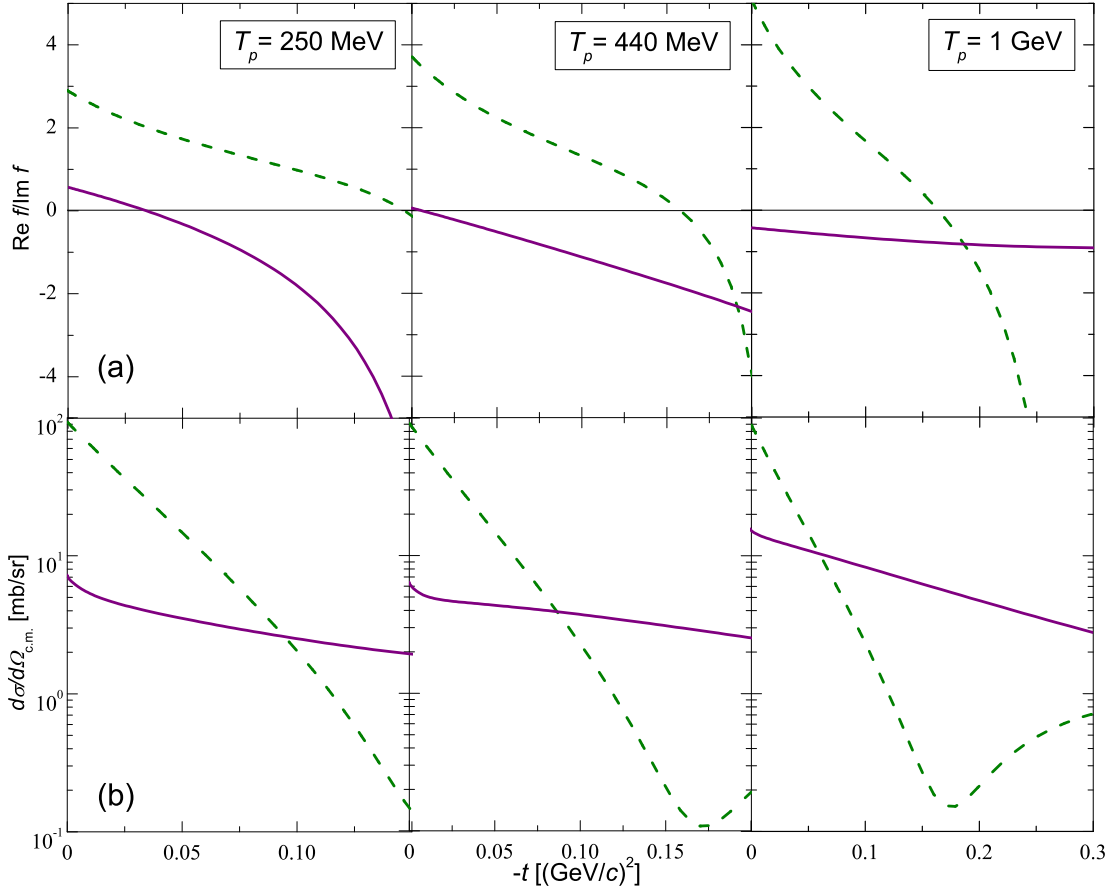}}
\end{center}
\caption{Ratio of real to imaginary parts
of the $NN$ spin-independent helicity amplitude (a) and $NN$
differential cross section (b) at different energies of the incident nucleon derived from the MT-III
potential model (dashed lines) with taken from PSA
\cite{SAID} for $np$ scattering (solid lines).} \label{nn-mt3}
\end{figure}

Our conclusion should be confronted with the results of the
previous work \cite{ELSTER} where a similar comparison was made between
the exact $3N$ Faddeev calculations and the conventional Glauber model
predictions for intermediate-energy $Nd$ scattering.
In that work, both theoretical approaches were based on the simple
central $NN$ potential MT-III (employed to calculate the input
$NN$ amplitude for the Glauber model), so the comparison between the
predictions was performed for differential and total cross
sections only. The authors \cite{ELSTER} found that in case
of the model $NN$ potential MT-III, the Glauber model results do
not reproduce the exact $3N$ calculation results for the differential
cross section at $T_N \simeq 200$ MeV and the predictions of both
approaches become more similar only at higher energies $T_N
\gtrsim 1$ GeV, as should be expected. Nevertheless, a fair
agreement between two approaches was found for the
single-scattering terms only, while the Glauber on-shell
double-scattering correction was shown to be insufficient in
comparison to the Faddeev second-order rescattering correction.
Thus, the general conclusion of Ref.~\cite{ELSTER} was that the
Glauber and fully converged Faddeev results \textit{do not
coincide} beyond the very forward angles (where the single
scattering dominates) even at the highest energy considered ($T_N
= 2$ GeV). However, when confronting both series of results one
should keep in mind that the model $NN$ potential MT-III does not
reproduce the empirical $NN$ scattering amplitudes at higher
partial waves $l \geq 1$, and thus does not reproduce the total
$NN$ amplitudes even at $T_N = 250$ MeV, see the
Fig.~\ref{nn-mt3}. It should be stressed that the Glauber approach
exploits essentially the characteristic features of just empirical
$NN$ amplitudes and with other types of the input $NN$ amplitudes
the contributions of neglected terms may become much higher. In
particular, the strong sensitivity of the Glauber model results
for $pd$ scattering (especially in the diffraction minimum) to the
ratio of real to imaginary parts of $NN$ amplitudes is well known
(see, for example, \cite{FDS}). Due to numerous inelastic
processes at $T_N
> 300$ MeV the realistic $NN$ potential has to have an imaginary part rising with
energy. This imaginary part of the $NN$ potential leads to an $NN$
scattering amplitude which has an enhanced imaginary part, while
the amplitude for the model MT-III potential has very small
imaginary part strongly decreasing with the rise of energy (see
Fig.~\ref{nn-mt3}, upper row).

A second but even more important
point is seen from the comparison of the model $NN$ differential
cross sections (for MT-III potential) with realistic ones (see
Fig.~\ref{nn-mt3}, lower row). The rates of falling for two types
of cross sections (as functions of momentum transfer squared) are
completely different, so the effective radius of the $NN$
interaction in the realistic case appears to be much shorter than that
for the model MT-III interaction. Indeed, the effective radius for
MT-III potential $r_{NN} \simeq 2$ fm or even more,\footnote{If to
define this radius as that value of $r_{NN}$ where the potential
can be practically neglected.} so that when analyzing
the double-scattering term with such a model input $NN$ potential and
keeping in mind that the average distance between two nucleons in
deuteron is around $4$ fm one can conclude that in this $NN$ model
the incident nucleon is moving through the target deuteron all the
time within the field of strong nuclear force. I.e., we cannot
consider the incident nucleon in this schematic model as moving
freely in between two successive collisions with the nucleons in
deuteron. In case of the realistic $NN$ interaction, the effective
range of $NN$ force gets much shorter as compared to the size of
deuteron (this is clearly seen from Fig.~\ref{nn-mt3}) and
thus the above assumption of the Glauber model for estimation of
the double-scattering term becomes quite valid.

An additional argument in favor of validity of the above Glauber
model assumption for the double-scattering term is the good
agreement between the diffraction model results and exact $3N$
calculations found for many observables, i.e., vector and tensor
analyzing powers as well as differential cross sections. In fact,
as is seen from Figs.~\ref{ds-ay-250}--\ref{ds-ay-440} the
agreement for spin observables is evident in the area where the
double-scattering term dominates. But this term includes a strong
interference between non-spin-flip, single-spin-flip and
double-spin-flip $NN$ helicity amplitudes, so that the behavior of the
intermediate propagator of the projectile (moving between two
successive collisions) should be of high importance to reproduce
all the considered spin observables.

(iii) To compare further the refined Glauber model results with
the experimental data and with exact Faddeev results (see Figs.~\ref{ds-ay-250}--\ref{ds-apd-1000}) one can observe that the
area where the diffraction model predictions begin to deviate
essentially from the exact $3N$ results almost coincides with that
where the latter begin to deviate from experimental data. In
other words, the refined Glauber model reproduces quite properly
the results of exact $3N$ calculations just in the region where
the Faddeev $3N$ framework reflects properly the underlying $3N$
dynamics, i.e. the dynamics which assumes the validity of the
conventional $2N$ and $3N$ force models and implies the nucleonic
and $\Delta$-isobar degrees of freedom only.

From this point of view, the deviation of exact Faddeev results
from the accurate experimental data on $pd$ scattering
\cite{SEK1} can imply that some hidden degrees of freedom
(e.g., dibaryonic, etc.) manifest themselves in large-angle $pd$
scattering. A strong additional argument in favor of just this
hypothesis follows from the fact that the above deviation gets
more and more serious when collision energy is rising. According
to some previous theoretical and experimental works (see, e.g.,
\cite{GUR,VIN}) the disagreements at $500$--$1000$ MeV may reach an
order of magnitude at large scattering angles.

(iv) The last, but not least, problem which can be posed by our
Glauber model calculations is related to the amazingly good
accuracy of the diffraction model at relatively low energies $T_p
\simeq 200$ MeV and at rather large scattering angles. To solve
this puzzle, one should recall that when considering scattering of
antiprotons by light nuclei the validity of the Glauber model was
found to begin at as low as $50$ MeV \cite{UZ}. The validity at
such a low energy is with no doubts tightly related to strong
absorption of antiprotons by the nuclear core, so that the central
nuclear area (where the nuclear density is still noticeable) is
seen by the incident antiproton as a large black disk, on which the
diffraction is observed in such experiments.

\begin{figure}
\begin{center}
\resizebox{0.33\columnwidth}{!}{\includegraphics{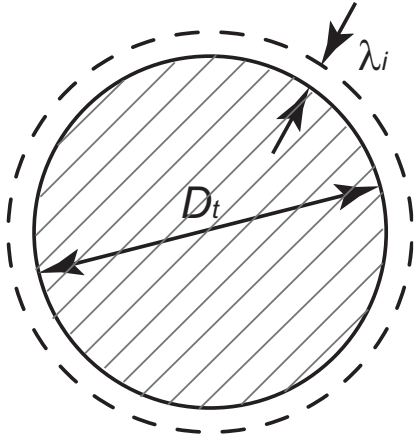}}
\end{center}
\caption{Illustration of optical diffraction in
high-energy $pd$ elastic scattering. The almost-black disk with radius
$D_t/2$ (hatched disk) surrounded by the wave zone of width
$\lambda_i$ (dashed line) represents the area inside the
loosely bound target where inelastic processes dominate. The
elastic scattering proceeds mainly in the ring of width
$\lambda_i$, so that $\lambda_i/D_t \ll 1$.} \label{disk}
\end{figure}

Rather similar physics is seen behind the intermediate- and
high-energy $pd$ scattering. Because the elastic $pd$ cross
section at these energies is rather small fraction of the total
cross section the dominating processes are just inelastic ones (at
least at small and middle impact parameter values), so that the
fast incident nucleon goes away from the elastic channel with high
probability when it is not very far from the loosely bound target
nucleus. Thus, the $pd$ elastic scattering at such high energies
can be viewed as a diffraction of the fast incident particle on the
edge of the large black disk, so that the diffraction process can be
described as a peripheral collision. This physical picture is
schematically represented by Fig.~\ref{disk}. Here the central
area (the hatched disk) with the radius $r_t = D_t/2$, with $D_t$
being the size of the deuteron ($D_t \simeq 4$ fm), shows the almost-black disk where the incident nucleons drop out of the elastic
scattering channel and undergo mainly inelastic scattering (an
``absorption'' from the entrance channel). So, the truly elastic
scattering happens mainly at the edge of the hatched disk inside a
ring (shown by the dashed line on Fig.~\ref{disk}) with the width
$\lambda_i$ (it corresponds to the wave zone in optical
diffraction). Thus, the ratio $\eta_{\sigma} =
\sigma_{el}/\sigma_{tot}$ of the elastic scattering cross section to
the total one can be roughly estimated as the ratio of areas
inside the ring and hatched inner disk, i.e., $\eta_r = \frac{2\pi
r_t \lambda_i}{\pi {r_t}^2} = 2 \lambda_i/r_t$. For the incident
energies $T_N \simeq 100$-$200$ MeV the nucleon wavelength
$\lambda_i \simeq 0.2$--$0.3$ fm, so that the ratio of the areas
$\eta_r = 2 \lambda_i/r_t \simeq 0.20$--$0.25$ which is in a good
qualitative agreement with the measured ratio $\eta_{\sigma} =
\sigma_{el}/\sigma_{tot} \simeq 0.15$--$0.20$. From this simple
picture one can understand clearly the reasons for a good
applicability of the Glauber diffraction model for the $pd$
elastic scattering even at energy $T_p \simeq 200$ MeV.

As for the observed validity of the Glauber model at rather large
transferred momenta ${\bf q}$, it is related basically to a
double-scattering term which dominates in the region beyond the
forward diffraction peak. So, the momentum ${\bf q}$ transferred
within the double scattering corresponds to ca. ${\bf q}/2$ for
each of single scatterings entering the double-scattering term.
Thus, it is very likely that although the validity of eikonal
approximation at ${\bf q}/2$ can be broken in a strict sense, the
degree of this breaking should increase rather slowly with the
rise of $q$.

(v) Finally, it would be very appropriate to discuss here some
possible reasons for observed disagreement between the results of
exact $3N$ calculations and experimental data for $pd$ cross
sections and especially for spin observables at large and backward
scattering angles. This topic can be important also to improve the
diffraction model description of the experimental data at larger
$t$-values.

The observation of $pd$ differential cross section and spin
analyzing powers at large scattering angles shows that starting
with incident energy $T_p \simeq 200$ MeV, the disagreements
between exact Faddeev calculations and respective experimental
data increase when the collision energy increases, and the
contribution of conventional $3N$ forces (induced by the
intermediate $\Delta$-isobar generation) does not help in reaching
the agreement~\cite{SEK1}. So, it seems that this observation makes it
meaningless to improve the formal aspects of Glauber model by
taking into account many other effects ignored in the present
formulation, e.g., off-shell corrections and relativistic effects such as
boosts, etc., because the majority of these effects have been already
included in the exact $3N$ calculations \cite{FC250} and likely do not
help reach a good agreement with the data at large
angles. It is important to stress also that the experimental
differential cross sections at large angles are typically
underestimated by the present-day theory. This fact and rise of
all disagreements with energy could imply that the theoretical
model does not include some essential degrees of freedom which
manifest themselves at rising energy stronger and stronger. One
can suppose that the most plausible candidature for this d.o.f.
ignored in all previous $3N$ calculations (and also in all
previous Glauber model results) are quark-meson (or dressed
dibaryon~\cite{AP05,DBM1}) d.o.f. Indeed, the dressed dibaryon describes
the situation when two nucleons overlap strongly their quark cores
(at $r_{NN} \lesssim 1$ fm). So, according to the modern dibaryon
concept~\cite{DBM1,DBM2}, this area corresponds to a strong attraction
between two quark cores due to an appearance of a strong scalar
field surrounding the unified six-quark system. In such a picture,
the incident nucleon scattered into large angles is feeling not
two well-separated nucleons in deuteron but one compact quark bag
which can survive, in sharp contrast to the loosely bound deuteron,
even at very large transferred momenta. Thus, if we assume for a
moment the existence of such a dressed dibaryon in deuteron with a
weight of about $2$--$3\%$~\cite{DBM1,DBM2}, it should be sufficient to
enhance strongly the backward scattering of intermediate- and
high-energy hadrons by deuteron. So, the straightforward
generalization of the Glauber model can be done also in this
direction.\footnote{In doing this, one should consider a direct
hadron-dibaryon interaction without basic approximations of the
diffraction model like eikonal, etc.}

\section{Summary}
\label{sum}

In this work, we presented the comparison between the predictions
for $pd$ elastic scattering observables given by the refined
Glauber model, exact Faddeev calculations, and experiments. As input for refined Glauber model, we used the fully realistic $NN$
helicity amplitudes which describe the $NN$ observables at
intermediate energies (at the level of accuracy of modern PSA) and the
high-precision model of the deuteron wave function. For the convenient
representation of the deuteron wave functions and $NN$
helicity amplitudes, we employed the special multi-Gaussian
expansion which allowed us to perform all the calculations fully
analytically. So, we calculated within the framework of the
refined diffraction model the differential cross sections and the
spin-dependent observables, i.e., the analyzing powers of proton and
deuteron. We found an amazingly good agreement between the results of our
refined Glauber model and exact Faddeev calculations up to
transferred momentum values where the exact $3N$ results begin to
deviate essentially from the experimental data. We discussed the
possible reasons for such surprising agreement, which extends to
rather low energies ($T_p \gtrsim 200$ MeV) and rather large
scattering angles.

Our general conclusion derived from the detailed comparisons with
exact $3N$ calculations and very numerous experimental data for
$pd$ analyzing powers and scattering cross sections can be
formulated as follows: the Glauber model (in its refined form
developed in the present work) turns out to be quite accurate
starting with relatively low energies for loosely bound target
nuclei as deuteron is. The refined diffraction model leads to
predictions (in a rather wide scope of its applicability) which
are, in general, in a similar agreement with experimental data as
the exact Faddeev calculations.

This conclusion should be valid not only for hadron scattering on
loosely bound nuclei such as $d$, ${}^6{\rm Li}$, etc., but also for
scattering of such hadrons as $\eta$, $K$ and other mesons on
arbitrary nuclei, i.e., in the case of strong absorption of an incident
wave by the nuclear core.

\acknowledgments The authors are very grateful to Prof. A. Faessler
for the nice hospitality in T\"{u}bingen University where part of this
work was done. We appreciate very much the partial financial
support from RFBR Grants Nos.~08-02-91959 and 07-02-00609 and the
DFG Grant No. 436 RUS 113/790/0-2.

\appendix
\section{Interrelations between $\bm{pd}$ and $\bm{NN}$ amplitudes in the refined
Glauber model} \label{A}

In this Appendix, we present the final formulas of the refined
Glauber model. These formulas relate the $pd$ invariant amplitudes
$A_1$--$A_{12}$ to the $NN$ invariant amplitudes
$A_i,B_i,C_i,C'_i,G_i,H_i$ ($i=n,p$) and deuteron form factors
$S_0^{(0)},S_0^{(2)},S_2^{(1)},S_2^{(2)}$ (for definitions, see
Eqs.~(\ref{ma}),(\ref{mi}), and (\ref{s00})). The general expression
for each $pd$ invariant amplitude finally takes the form
\begin{equation*}
 A_j(q) = [A_j^{(s)}(q) + A_j^{(d)}(q) + A_j^{(c)}(q)] + [n
\leftrightarrow p],
\end{equation*}
\begin{equation}
\label{aj} A_j^{(d)}(q) = {\textstyle \frac{i}{2 \pi^{3/2}}
\int} d^{2}q' \mathcal{A}_j^{(d)}({\bf q},{\bf q'}), \quad
A_j^{(c)}(q) = {\textstyle \frac{i}{2 \pi^{3/2}} \int} d^{2}q'
\bigl(\mathcal{A}_j^{(d)}({\bf q},{\bf q'}) -
\mathcal{A}_j^{(c)}({\bf q},{\bf q'})\bigr),
\end{equation}
where $j=\overline{1,12}$, and ``$+[n \leftrightarrow p]$'' denotes an
addition of the preceding expression (in square brackets) with the
neutron and proton indices interchanged throughout. The formulas
for the quantities $A_j^{(s)}, \mathcal{A}_j^{(d)}$ and
$\mathcal{A}_j^{(c)}$ can be found in Tables~\ref{t1},\ref{t2}, and
\ref{t3}, respectively. It is implied there that in
single-scattering amplitudes $A_j^{(s)}$ the deuteron formfactors
are functions of ${\bf q}/2$ and invariant $NN$ amplitudes are
functions of ${\bf q}$. In the quantities $\mathcal{A}_j^{(d)}$
and $\mathcal{A}_j^{(c)}$, entering the double-scattering and
double-charge-exchange amplitudes ($A_j^{(d)}$ and $A_j^{(c)}$),
the deuteron formfactors are functions of ${\bf q'}$ while in
products of $NN$ amplitudes the first one depends on ${\bf q_2}$
and the second one depends on ${\bf q_1}$, e.g. $A_n A_p \equiv
A_n({\bf q_2}) A_p({\bf q_1})$.

In the derivation of $\mathcal{A}_j^{(d)}$ and $\mathcal{A}_j^{(c)}$,
we used the following approximate interrelations between the unit
vectors:
\begin{equation}
 \hat{k} \approx \hat{k}_1 \approx \hat{k}_2, \\
\end{equation}
\begin{equation}
 \label{nq}
 \hat{n}_i \hat{q}_l \approx \hat{q}_i \!\times\! \hat{q}_l, \ \
 \hat{n}_i \hat{n}_l \approx \hat{q}_i \hat{q}_l.
\end{equation}
Here $i,l \in \{0,1,2\}$, and we define $\hat{k}_0 \equiv \hat{k},
\hat{q}_0 \equiv \hat{q}, \hat{n}_0 \equiv \hat{n}$, and introduce the unit vectors $\hat{k}_i,\hat{q}_i,\hat{n}_i, \
i=1,2$ for two individual collisions in the double scattering. The
above interrelations are valid within the eikonal approximation.

\begin{table}[!ht]
\caption{Formulas for the single-scattering amplitudes $A_j^{(s)},
\ j=\overline{1,12}$ (see Eq.~(\ref{aj})).}
\bigskip
\begin{center}
\resizebox{0.4\textwidth}{!}{\begin{tabular}{|rcl|} \hline
  $A_1^{(s)}$ &$=$& $\bigl(S_0 + \sqrt{2} S_2\bigr) A_n$ \\
\hline
  $A_2^{(s)}$ &$=$& $\bigl(S_0^{(0)} + \sqrt{2} S_2^{(1)}\bigr) C_n$ \\
\hline
  $A_3^{(s)}$ &$=$& $-{\textstyle\frac{3}{\sqrt{2}}} S_2 A_n$ \\
\hline
  $A_4^{(s)}$ &$=$& $-{\textstyle\frac{3}{\sqrt{2}}} S_2^{(1)} C_n$ \\
\hline
  $A_5^{(s)}$ &$=$& $0$ \\
\hline
  $A_6^{(s)}$ &$=$& $0$ \\
\hline
  $A_7^{(s)}$ &$=$& $\Bigl(S_0^{(0)} + {\textstyle\frac{1}{2\sqrt{2}}} S_2^{(1)}\Bigr) B_n$ \\
\hline
  $A_8^{(s)}$ &$=$& $0$ \\
\hline
  $A_9^{(s)}$ &$=$& $\Bigl(S_0^{(0)} + {\textstyle\frac{1}{2\sqrt{2}}} S_2^{(1)}\Bigr) C'_n$ \\
\hline
  $A_{10}^{(s)}$ &$=$& $\Bigl(S_0^{(0)} + {\textstyle\frac{1}{2\sqrt{2}}} S_2^{(1)}\Bigr) (G_n - H_n)$ \\
\hline
  $A_{11}^{(s)}$ &$=$& $\Bigl(S_0^{(0)} - {\textstyle\frac{1}{\sqrt{2}}} S_2^{(1)}\Bigr) (G_n + H_n)$ \\
\hline
  $A_{12}^{(s)}$ &$=$& $0$ \\
\hline
\end{tabular}}
\end{center} \label{t1}
\end{table}
\clearpage
\begin{table}[!ht]
\caption{Formulas for the quantities $\mathcal{A}_j^{(d)}$
entering the expressions for double-scattering amplitudes
$A_j^{(d)}, \ j=\overline{1,12}$ (see Eq.~(\ref{aj})).}
\bigskip \resizebox{1.0\textwidth}{!}{
\begin{tabular}{|rcl|} \hline
   $\mathcal{A}_1^{(d)}$ &$=$& ${\textstyle\frac{1}{2}}S_0^{(0)} \Bigl( A_n A_p + 3B_n B_p + \bigl(C_n C_p -
C'_n C'_p \bigr)\hat{q}_2\hat{q}_1 - 2 G_n G_p - 2H_n H_p
\bigl((\hat{q}_2\hat{q}_1)^2 - (\hat{q}_2
\!\times\! \hat{q}_1)^2 \bigr)\Bigr) + $ \\
& & $ + {\textstyle\frac{1}{2}} \bigl( S_0^{(2)} + \sqrt{2} S_2 \bigr) A_n A_p $ \\
\hline
   $\mathcal{A}_2^{(d)}$ &$=$& $S_0^{(0)} \Bigl(A_n C_p \,\hat{q}\hat{q}_1 - C'_n
G_p \,\hat{q}\hat{q}_2 + C'_n H_p \bigl((\hat{q}_2
\hat{q}_1)(\hat{q}\hat{q}_1) - (\hat{q}_2 \!\times\!
\hat{q}_1)(\hat{q} \!\times\! \hat{q}_1) \bigr)\Bigr) +
$ \\
& & $ + \sqrt{2} S_2^{(1)} A_n C_p \,\hat{q}\hat{q}_1 $ \\
\hline
   $\mathcal{A}_3^{(d)}$ &$=$& $S_0^{(0)} \Bigl(C'_n C'_p(\hat{q}
\!\times\! \hat{q}_2)(\hat{q} \!\times\! \hat{q}_1) - B_n B_p +
G_n G_p + H_n H_p \bigl((\hat{q}_2\hat{q}_1)^2
-(\hat{q}_2 \!\times\! \hat{q}_1)^2 \bigr) + $ \\
& & $ + 2 G_n H_p \bigl((\hat{q}\hat{q}_1)^2 -(\hat{q} \!\times\!
\hat{q}_1)^2 \bigr)\Bigr) - {\textstyle\frac{3}{2\sqrt{2}}} S_2
A_n A_p (\hat{q}\hat{q}')^2  $ \\
\hline
   $\mathcal{A}_4^{(d)}$ &$=$& $-4S_0^{(0)} C'_n H_p(\hat{q} \!\times\! \hat{q}_2)
(\hat{q} \!\times\! \hat{q}_1)(\hat{q}\hat{q}_1) -
{\textstyle\frac{3}{\sqrt{2}}} S_2^{(1)} A_n C_p \,\hat{q}\hat{q}_1(\hat{q}\hat{q}')^2 $ \\
\hline
   $\mathcal{A}_5^{(d)}$ &$=$& $S_0^{(0)} \Bigl(C'_n C'_p(\hat{q}
\hat{q}_2)(\hat{q}\hat{q}_1) - B_n B_p + G_n G_p + H_n H_p
\bigl((\hat{q}_2\hat{q}_1)^2 -(\hat{q}_2
\!\times\! \hat{q}_1)^2 \bigr) - $ \\
& & $ - 2 G_n H_p \bigl((\hat{q}\hat{q}_1)^2 -(\hat{q} \!\times\!
\hat{q}_1)^2 \bigr)\Bigr) - {\textstyle\frac{3}{2\sqrt{2}}} S_2
A_n A_p (\hat{q} \!\times\! \hat{q}')^2  $ \\
\hline
   $\mathcal{A}_6^{(d)}$ &$=$& $2S_0^{(0)} \Bigl( C'_n G_p - C'_n H_p\bigl((\hat{q}\hat{q}_1)^2 - (\hat{q}
\!\times\! \hat{q}_1)^2\bigr)\Bigr)\,\hat{q}\hat{q}_2 -
{\textstyle\frac{3}{\sqrt{2}}} S_2^{(1)} A_n C_p
\,\hat{q}\hat{q}_1(\hat{q}
\!\times\! \hat{q}')^2  $ \\
\hline
   $\mathcal{A}_7^{(d)}$ &$=$&  $\bigl(S_0^{(0)} +
{\textstyle\frac{1}{2\sqrt{2}}} S_2^{(1)}\bigr)A_n B_p  $ \\
\hline
   $\mathcal{A}_8^{(d)}$ &$=$&  $S_0^{(0)} \Bigl(C'_n G_p \,\hat{q}\hat{q}_2 + C'_n H_p \bigl(\hat{q}\hat{q}_2 \bigl((\hat{q}\hat{q}_1)^2 -(\hat{q}
\!\times\! \hat{q}_1)^2\bigr) - 2(\hat{q} \!\times\!
\hat{q}_2)(\hat{q} \!\times\! \hat{q}_1)(\hat{q}\hat{q}_1)\bigr)\Bigr) + $ \\
 & & $ + {\textstyle\frac{3}{\sqrt{2}}} S_2^{(1)} A_n C_p(\hat{q} \!\times\! \hat{q}_1)(\hat{q} \!\times\!
\hat{q}')(\hat{q}\hat{q}') $ \\
\hline
   $\mathcal{A}_9^{(d)}$ &$=$& $S_0^{(0)} \Bigl(A_n C'_p \,\hat{q}\hat{q}_1 + C_n
G_p \,\hat{q}\hat{q}_2 - C_n H_p
\bigl((\hat{q}_2\hat{q}_1)(\hat{q}\hat{q}_1) - (\hat{q}_2
\!\times\! \hat{q}_1)(\hat{q} \!\times\! \hat{q}_1) \bigr)\Bigr) + $ \\
 & & $ + {\textstyle\frac{1}{2\sqrt{2}}} S_2^{(1)} A_n C'_p \bigl(\hat{q}\hat{q}_1 -
3(\hat{q}_1 \!\times\! \hat{q}')(\hat{q} \!\times\!
\hat{q}')\bigr) $ \\
\hline
   $\mathcal{A}_{10}^{(d)}$ &$=$& $S_0^{(0)} \Bigl(C_n C'_p(\hat{q}\hat{q}_2)
(\hat{q}\hat{q}_1) + A_n G_p - A_n H_p \bigl((\hat{q}\hat{q}_1)^2
- (\hat{q} \!\times\! \hat{q}_1)^2 \bigr)\Bigr) +
{\textstyle\frac{1}{2\sqrt{2}}} S_2^{(1)} \Bigl(A_n G_p \times $ \\
& & $ \times \bigl(1-3(\hat{q} \!\times\!
 \hat{q}')^2\bigr) - A_n H_p \bigl[(\hat{q}\hat{q}_1)^2 - (\hat{q} \!\times\! \hat{q}_1)^2 - 3(\hat{q} \!\times\!
 \hat{q}') \bigl((\hat{q}\hat{q}_1)(\hat{q}_1 \!\times\! \hat{q}') - (\hat{q} \!\times\! \hat{q}_1)(\hat{q}_1\hat{q}')\bigr)
 \bigr]\Bigr) $ \\
\hline
   $\mathcal{A}_{11}^{(d)}$ &$=$& $S_0^{(0)} \Bigl(C_n C'_p(\hat{q} \!\times\! \hat{q}_2)
(\hat{q}\!\times\! \hat{q}_1) + A_n G_p + A_n H_p
\bigl((\hat{q}\hat{q}_1)^2 -(\hat{q} \!\times\! \hat{q}_1)^2
\bigr)\Bigr) + {\textstyle\frac{1}{2\sqrt{2}}} S_2^{(1)} \Bigl(A_n
G_p \times $ \\
& & $ \times \bigl(1-3(\hat{q}\hat{q}')^2\bigr) + A_n H_p
\bigl[(\hat{q}\hat{q}_1)^2 - (\hat{q} \!\times\! \hat{q}_1)^2 -
3\hat{q}\hat{q}'\bigl((\hat{q}\hat{q}_1)(\hat{q}_1\hat{q}') -
(\hat{q} \!\times\! \hat{q}_1)(\hat{q}_1 \!\times\!
\hat{q}')\bigr)
 \bigr]\Bigr) $ \\
\hline $\mathcal{A}_{12}^{(d)}$ &$=$& $S_0^{(0)} C'_n B_p
\,\hat{q}\hat{q}_2
 $ \\
\hline
\end{tabular}}
\label{t2}
\end{table}
\clearpage
\begin{table}[!ht]
\caption{Formulas for the quantities $\mathcal{A}_j^{(c)}$
entering the expressions for double-charge-exchange amplitudes
$A_j^{(c)}, \ j=S_2^{(1)}$ (see Eq.~(\ref{aj})).}
\bigskip \resizebox{1.0\textwidth}{!}{
\begin{tabular}{|rcl|}
\hline
    $\mathcal{A}_1^{(c)}$ &$=$& ${\textstyle\frac{1}{2}}S_0^{(0)} \Bigl( A_n A_n + B_n B_n +
\bigl(C_n C_n + C'_n C'_n \bigr)\hat{q}_2\hat{q}_1 + 2 G_n G_n +
2H_n H_n \bigl((\hat{q}_2\hat{q}_1)^2 -
(\hat{q}_2 \!\times\! \hat{q}_1)^2 \bigr)\Bigr) + $ \\
 & & $ + {\textstyle\frac{1}{2}} \bigl(S_0^{(2)} + \sqrt{2} S_2 \bigr) A_n A_n $ \\
\hline
   $\mathcal{A}_2^{(c)}$ &$=$& $S_0^{(0)} \Bigl(A_n C_n \,\hat{q}\hat{q}_1 + C'_n
G_n \,\hat{q}\hat{q}_2 - C'_n H_n
\bigl((\hat{q}_2\hat{q}_1)(\hat{q}\hat{q}_1) - (\hat{q}_2
\!\times\! \hat{q}_1)(\hat{q} \!\times\! \hat{q}_1) \bigr)\Bigr) + $ \\
& & $ + \sqrt{2} S_2^{(1)} A_n C_n \,\hat{q}\hat{q}_1 $ \\
\hline
   $\mathcal{A}_3^{(c)}$ &$=$& $-{\textstyle\frac{3}{2\sqrt{2}}} S_2
A_n A_n (\hat{q}\hat{q}')^2 $ \\
\hline
   $\mathcal{A}_4^{(c)}$ &$=$& $-{\textstyle\frac{3}{\sqrt{2}}} S_2^{(1)} A_n C_n \,\hat{q}\hat{q}_1(\hat{q}\hat{q}')^2 $ \\
\hline
   $\mathcal{A}_5^{(c)}$ &$=$& $-{\textstyle\frac{3}{2\sqrt{2}}} S_2
A_n A_n (\hat{q} \!\times\! \hat{q}')^2 $ \\
\hline
   $\mathcal{A}_6^{(c)}$ &$=$& $-{\textstyle\frac{3}{\sqrt{2}}} S_2^{(1)} A_n C_n \,\hat{q}\hat{q}_1(\hat{q} \!\times\! \hat{q}')^2 $ \\
\hline
   $\mathcal{A}_7^{(c)}$ &$=$& $S_0^{(0)} \Bigl( A_n B_n - G_n G_n + H_n H_n
\bigl((\hat{q}_2\hat{q}_1)^2 - (\hat{q}_2 \!\times\!
\hat{q}_1)^2 \bigr)\Bigr) + {\textstyle\frac{1}{2\sqrt{2}}} S_2^{(1)} A_n B_n $ \\
\hline
   $\mathcal{A}_8^{(c)}$ &$=$& ${\textstyle\frac{3}{\sqrt{2}}} S_2^{(1)} A_n C_n(\hat{q} \!\times\!
\hat{q}_1)(\hat{q} \!\times\! \hat{q}')(\hat{q}\hat{q}') $ \\
\hline
   $\mathcal{A}_9^{(c)}$ &$=$& $S_0^{(0)} \Bigl(A_n C'_n \,\hat{q}\hat{q}_1 + C_n
G_n \,\hat{q}\hat{q}_2 - C_n H_n
\bigl((\hat{q}_2\hat{q}_1)(\hat{q}\hat{q}_1) - (\hat{q}_2
\!\times\! \hat{q}_1)(\hat{q} \!\times\! \hat{q}_1) \bigr)\Bigr) + $ \\
 & & $ + {\textstyle\frac{1}{2\sqrt{2}}} S_2^{(1)} A_n C'_n \bigl(\hat{q}\hat{q}_1 -
3(\hat{q}_1 \!\times\! \hat{q}')(\hat{q} \!\times\!
\hat{q}')\bigr) $ \\
\hline
   $\mathcal{A}_{10}^{(c)}$ &$=$& $S_0^{(0)} \Bigl(C_n C'_n(\hat{q}\hat{q}_2)
(\hat{q}\hat{q}_1) + \bigl(A_n - B_n \bigr)G_n - \bigl(A_n + B_n
\bigr) H_n \bigl((\hat{q}\hat{q}_1)^2 - (\hat{q} \!\times\!
\hat{q}_1)^2
\bigr)\Bigr) + $ \\
 & & $ + {\textstyle\frac{1}{2\sqrt{2}}} S_2^{(1)} \Bigl(A_n G_n\bigl(1-3(\hat{q} \!\times\!
 \hat{q}')^2\bigr) - A_n H_n \bigl[(\hat{q}\hat{q}_1)^2 - (\hat{q} \!\times\! \hat{q}_1)^2 - 3(\hat{q} \!\times\!
 \hat{q}') \bigl((\hat{q}\hat{q}_1)(\hat{q}_1 \!\times\! \hat{q}') - $ \\
 & & $ - (\hat{q} \!\times\! \hat{q}_1)(\hat{q}_1\hat{q}')\bigr)
 \bigr]\Bigr) $ \\
\hline
   $\mathcal{A}_{11}^{(c)}$ &$=$& $S_0^{(0)} \Bigl(C_n C'_n(\hat{q} \!\times\! \hat{q}_2)
(\hat{q}\!\times\! \hat{q}_1) + \bigl(A_n - B_n \bigr) G_n +
\bigl(A_n + B_n \bigr) H_n\bigl((\hat{q}\hat{q}_1)^2 -(\hat{q}
\!\times\! \hat{q}_1)^2 \bigr)\Bigr) + $ \\
 & & $ + {\textstyle\frac{1}{2\sqrt{2}}} S_2^{(1)} \Bigl(A_n G_n\bigl(1-3(\hat{q}\hat{q}')^2\bigr) +
 A_n H_n \bigl[(\hat{q}\hat{q}_1)^2 - (\hat{q} \!\times\! \hat{q}_1)^2 - 3\hat{q}\hat{q}'\bigl((\hat{q}\hat{q}_1)(\hat{q}_1\hat{q}') - $ \\
 & & $ - (\hat{q} \!\times\! \hat{q}_1)(\hat{q}_1 \!\times\! \hat{q}')\bigr)
 \bigr]\Bigr) $ \\
\hline
   $\mathcal{A}_{12}^{(c)}$ &$=$& $0$ \\
\hline
\end{tabular}}
\label{t3}
\end{table}

\section{Analytical calculation of the integrals in double-scattering amplitudes}
\label{B}

With the special parametrization of the input $NN$ helicity amplitudes and deuteron wave
functions presented in Sec.~\ref{param}, the ${\bf q'}$ integration in the double-scattering and
double-charge-exchange amplitudes $A_j^{(d)}$ and $A_j^{(c)}, \
j=\overline{1,12}$ (see Eq.~(\ref{aj})) can be performed fully analytically.
In particular, the scalar and vector products of the unit vectors
$\hat{q},\hat{q}',\hat{q}_1,\hat{q}_2$ (appearing in Tables~\ref{t2} and \ref{t3})
are expressed through their
magnitudes $q,q',q_1,q_2$ and the angle $\varphi$ between
$\hat{q}$ and $\hat{q}'$ as
\begin{gather}
 \hat{q} \hat{q}' = \cos\varphi, \ \ \hat{q} \!\times\!
\hat{q}' = \sin\varphi, \nonumber \\
 \hat{q} \hat{q}_1 = (q/2
- q'\cos\varphi)/q_1,  \ \ \hat{q} \!\times\! \hat{q}_1 =
-q'\sin\varphi/q_1, \nonumber \\
 \hat{q} \hat{q}_2) = (q/2 +
q'\cos\varphi)/q_2,  \ \ \hat{q} \!\times\! \hat{q}_2 =
q'\sin\varphi/q_2, \nonumber \\
 \hat{q}_1 \hat{q}' = (-q' +
(q/2)\cos\varphi)/q_1,  \ \ \hat{q}_1 \!\times\! \hat{q}' =
(q/2)\sin\varphi/q_1, \nonumber \\
 \hat{q}_2 \hat{q}' = (q' +
(q/2)\cos\varphi)/q_2,  \ \ \hat{q}_2 \!\times\! \hat{q}' =
(q/2)\sin\varphi/q_2, \nonumber \\
 \hat{q}_2 \hat{q}_1 = (q^2/4 - q'^2)/(q_2q_1),  \ \
\hat{q}_2 \!\times\! \hat{q}_1 = -qq'\sin\varphi/(q_2q_1),
\nonumber \\
 q_1^2 = q^2/4 + q'^2 - qq'\cos\varphi, \ \ q_2^2 = q^2/4 + q'^2
+ qq'\cos\varphi.
\label{vec}
\end{gather}
When multiplying these products by the $NN$ amplitudes, the
magnitudes $q_1$ and $q_2$ in denominators are exactly canceled
with the factors which represent the behavior of the $NN$
amplitudes near the origin (see Eq.~(\ref{nug})). Thus, making use
of the expansions for $NN$ amplitudes (Eq.~(\ref{nug})) and deuteron
form factors (Eq.(\ref{sg})) and the Eq.~(\ref{vec}), all integrals can be reduced to the standard form
\begin{gather}
J_{mn}(\alpha,\beta;q) \equiv {\textstyle \int \limits_0^{\infty}}
dq' {\textstyle \int \limits_0^{2 \pi}} d\varphi q'^n e^{-\alpha
q'^2 + \beta q q' \cos\varphi} \cos(m\varphi) = \nonumber \\
= {\textstyle \frac{\pi \Gamma((n+m+1)/2) \beta^m q^m}{2^m
\Gamma(m+1) \alpha^{(n+m+1)/2}}} {\rm {}_1F_1}((n+m+1)/2, m+1, \beta^2
q^2/(4\alpha)),
\end{gather}
throughout, where $n \geq 0$, $m \geq 0$ are integer numbers, and $\alpha = A1+A2+1/(4\lambda)$, $\beta = A1-A2$ are the combinations of nonlinear Gaussian
parameters ($\lambda$ comes from
the deuteron form factors, while $A1$ and $A2$ are related to the $NN$ amplitudes depending on
${\bf q1}$ and ${\bf q2}$, respectively). The confluent hypergeometric function ${\rm {}_1F_1}$ in our case has positive integer
numbers in its first two arguments, so it can be expressed
through simple Gaussians and polynomials in $q$. As a result, one obtains the fully
analytical expressions for all $pd$ invariant amplitudes
$A_1$--$A_{12}$ in terms of input Gaussian parameters of $NN$
helicity amplitudes and deuteron wave functions.

\end{document}